\documentclass[twocolumn,showpacs,preprintnumbers,amsmath,amssymb]{revtex4}
\usepackage{graphicx}% Include figure files
\usepackage{dcolumn}% Align table columns on decimal point
\usepackage{bm,color}% bold math
\def\be{\begin{equation}}
\def\ee{\end{equation}}
\def\bea{\begin{eqnarray}}
\def\eea{\end{eqnarray}}

\def\md{\mathrm{d}}

\def\s{\sigma}

\begin{document}

%\preprint{}

\title{Next-to-Leading-Order study on the associate production of $J/\psi+\gamma$ at the LHC}% Force line breaks with \\



\author{Rong Li$^{1,3}$ and Jian-Xiong Wang$^{2,3}$}
\address{$^1$Department of Applied Physics, Xi'an Jiaotong University, Xi'an 710049,
China.\\
$^2$Institute of High Energy Physics, Chinese Academy of Sciences,
P.O. Box 918(4), Beijing, 100049, China.\\
$^3$Theoretical Physics Center for Science Facilities, CAS, Beijing,
100049, China.}

%\thanks{}
%\affiliation{}


\date{\today}

\begin{abstract}
The associate $J/\psi+\gamma$ production at the LHC is studied
completely at next-to-leading-order (NLO) within the framework of
nonrelativistic QCD. By using three sets of color-octet
long-distance matrix elements (LDMEs) obtained in previous prompt
$J/\psi$ studies, we find that only one of them can result in a
positive transverse momentum ($p_t$) distribution of $J/\psi$
production rate at large $p_t$ region. Based on reasonable
consideration to cut down background, our estimation is measurable
upto $p_t=50$GeV with present data sample collected at $8$TeV LHC.
All the color-octet LDMEs in $J/\psi$ production could be fixed
sensitively by including this proposed measurement and our
calculation, and then confident conclusion on $J/\psi$ polarization puzzle
could be achieved.

%In comparison with
%previous color-singlet results at QCD NLO, the $p_t$ distribution of
%the $J/\psi$ production rate is enhanced about two orders of
%magnitude in the larger $p_t$ region and the polarization of
%$J/\psi$ changes from the longitudinal state in the color-singlet
%channel to the transverse one.
%Because this process is sensitive to
%the $\langle O^{J/\psi}(^3P_0^8)\rangle$, which is different from
%the inclusive production of $J/\psi$, these results provide a new
%supplement to investigate the production mechanism of $J/\psi$ and
%show that further study on the universality of the NRQCD LDMEs is
%needed.
\end{abstract}

%\pacs{14.40.Pq, 12.38.Bx, 13.85.Ni, 13.87.Fh}
%\keywords{}

\maketitle


Since the discovery of heavy quarkonium in the 1970s, the study on
production and decay of $J/\psi$ and $\Upsilon$, plays an important
role in the research on the perturbative and nonperturbative aspects
of QCD. In 1995, a new factorization framework, nonrelativistic QCD
(NRQCD), had been proposed to study the production and decay of
heavy quarkonium~\cite{Bodwin:1994jh}. It overcomes some
shortcomings in the prevalent color-single
model(CSM)~\cite{Einhorn:1975ua} and makes the CSM be a part of it.
By extracting the NRQCD long-distance matrix elements (LDMEs) from
the matching between theoretical prediction and experimental data,
the NRQCD calculation had given a well description on the transverse
momentum distribution ($p_t$) of heavy quarkonium production at
hadron colliders at leading order (LO)~\cite{Cho:1995vh}. The
phenomenological applications of the NRQCD have been investigated
extensively~\cite{Brambilla:2004wf}. But the polarization of heavy
quarkonium hadroproduction had been an open question for more than
ten years.

The next-to-leading order (NLO) QCD correction to the $J/\psi$
inclusive hadroproduction in the CSM significantly enhanced the
$p_t$ distribution~\cite{Campbell:2007ws} and changed the
polarization from transverse to longitudinal~\cite{Gong:2008sn}. The
later study including the contribution from color-octet model (COM)
parts($^1S_0^8$ and $^3S_1^8$) at QCD NLO still can not give the
satisfied prediction on the polarization of
$J/\psi$~\cite{Gong:2008ft}. In Ref.~\cite{Ma:2010vd}, the study on
P-wave charmonium hadroproduction with the feeddown from $\chi_{cJ}$
to $J/\psi$ had been obtained at QCD NLO.  Soon after the $p_t$
distribution of $J/\psi$ production rate at full NLO QCD had been
given by two groups~\cite{Butenschoen:2010rq,Ma:2010yw}. Then after
two years, the $p_t$ distributions on polarization for direct
$J/\psi$ hadroproduction at full NLO QCD were presented by two
group~\cite{Butenschoen:2012px, Chao:2012iv}. A few months later, a
complete study~\cite{Gong:2012ug} on the polarization of prompt
$J/\psi$ hadroproduction with the feeddown contribution from
$\chi_{cJ}$ included was given, and it present the first result at
QCD NLO which can be compared with the experimental measurements
directly since all the polarization measurements are for prompt
$J/\psi$. The recent measurements at the LHC by
CMS~\cite{Chatrchyan:2013cla} and LHCb~\cite{Aaij:2013nlm} show
disagreement with the prediction~\cite{Gong:2012ug}. However, the
results from three groups~\cite{Butenschoen:2012px,
Chao:2012iv,Gong:2012ug} imply that the $\chi^2$ fit to extract the
LDMEs of $J/\psi$ production is very sensitive to input condition
due to the approximately linear correlation of the three color-octet
contribution parts. Therefore, no solid conclusion on the
polarization of $J/\psi$ could be achieved, and other relevant
reliable perturbative prediction, which is experimental mensurable
and can break the linear correlation in previous fit, is expected.
Is the study on associate $J/\psi+\gamma$ hadroproduction at full
QCD NLO a good candidate? Our study clearly indicates that it is a
very good candidate with present data sample collected at $8$TeV
LHC.

In another way, for $\Upsilon$ hadroproduction,
there are studies on the $p_t$ distribution of yield and
polarization for the CS channel at QCD
NLO~\cite{Campbell:2007ws,Gong:2008sn} and at the partial
next-to-next-to-leading order~\cite{Artoisenet:2008fc}. The NLO QCD
correction to $p_t$ distribution of the yield and polarization for
$\Upsilon(1S,3S)$ via S-wave CO states is presented in
Ref.~\cite{Gong:2010bk}, and the NLO QCD correction to $p_t$
distribution of the yield for $\Upsilon(1S)$ via all the CO states
is presented in Ref.~\cite{Wang:2012is}. The
complete NLO study on polarization of prompt $\Upsilon$
hadroproduction has been achieved in Ref.~\cite{Gong:2013qka}, which can
explain the recent measurement on the polarization of
$\Upsilon(1S,2S,3S)$ at the LHC by CMS
collaboration~\cite{Chatrchyan:2012woa}.

In addition to the study on important inclusive heavy quarkonium hadroproduction,
study on the associate hadroproduction of heavy
quarkonium and photon (or $W^{\pm}$,$Z^0$ bosons) was proposed
as a supplemental channel to probe the gluon content in the
proton~\cite{Drees:1991ig} or to investigate the production
mechanism of heavy quarkonium~\cite{Kim:1994bm}.
The NLO QCD correction to $J/\psi+W^{\pm}(Z^0)$ had been calculated in Ref.~\cite{Li:2010hc}.
Our study~\cite{Li:2008ym} shows the NLO QCD correction to the associate $J/\psi+\gamma$
hadroproduction in the CSM enhanced the $p_t$ distribution largely in the high momentum
region and changed the polarization from transverse to longitudinal.
The relevant study~\cite{Lansberg:2009db} can reproduce our results in a partial NLO calculation.
To obtain an experimental measurable observable at full QCD NLO,
we present the study on associate $J/\psi+\gamma$ hadroproduction at the NLO
with full COM contribution in this work.

In the NRQCD framework the inclusive production of $J/\psi+\gamma$
can be factorized as \bea &&\sigma(p+\bar{p} \to J/\psi+\gamma+X)
=\sum_{i,j} \int dx_1dx_2 \\ &\times& G_p^i(x_1) G_{\bar{p}}^j(x_2)
\hat{\sigma}(ij\to (Q\bar{Q})_n+\gamma+X)\langle
O^{J/\psi}_n\rangle. \nonumber \label{eqn:factorization} \eea Here
$G^{i(j)}_{p(\bar{p})}$ are the parton distribution functions(PDFs),
$\hat{\sigma}$ presents the parton level cross section, and $\langle
O^{J/\psi}_n\rangle$ are the LDMEs.
The relevant parton level processes are listed as following \bea
g+g \to Q\bar{Q}\lbrack ^3S_1^1,^1S_0^8,^3S_1^8,^3P_J^8\rbrack +\gamma, \label{eqn:vcom}  \\
g+g \to Q\bar{Q}\lbrack ^3S_1^1,^1S_0^8,^3S_1^8,^3P_J^8\rbrack + \gamma +g, \label{eqn:r1com}  \\
q+\bar{q} \to Q\bar{Q}\lbrack ^3S_1^1,^1S_0^8,^3S_1^8,^3P_J^8\rbrack +\gamma,\label{eqn:r2com}\\
q+\bar{q} \to Q\bar{Q}\lbrack ^1S_0^8,^3S_1^8,^3P_J^8\rbrack +\gamma +g \label{eqn:r3com},\\
q(\bar{q})+g \to Q\bar{Q}\lbrack
^3S_1^1,^1S_0^8,^3S_1^8,^3P_J^8\rbrack + \gamma + q(\bar{q}).
\label{eqn:r4com} \eea
%In the color-singlet processes which have
%$Q\bar{Q}\lbrack^3S_1^1\rbrack$ as final states, the heavy quark
%anti-quark pair is produced in the short distance parts with the
%same quantum numbers($^3S_1^1$) as $J/\psi$ meson. The
%$Q\bar{Q}\lbrack^3S_1^1\rbrack$ states then have the definite
%probability to form $J/\psi$ meson that can be numerically related
%to the wave function at origin of $J/\psi$. All the above processes
%can have color-octet final states $Q\bar{Q}\lbrack
%^1S_0^8,^3S_1^8,^3P_J^8\rbrack$ produced in the short distance parts
%and then evolved into the $J/\psi$ meson by radiating soft gluons to
%adjust their color and angular states.

In addition to the $p_t({J/\psi})$ distribution of the $J/\psi+\gamma$
hadroproduction at QCD NLO, the related polarization observable $\alpha$ of $J/\psi$ is
also studied. $\alpha$ is measured by
using the angular distribution of the decayed lepton pair in
helicity frame and related to the spin density matrix of $J/\psi$ as
\bea \alpha(p_t)=\frac{{\md\s_{11}}/{\md p_t}- {\md\s_{00}}/{\md
p_t}}{{\md\s_{11}}/{\md p_t}+{\md\s_{00}}/{\md p_t}}.
\eea
Here the "1" and "0" mean the transverse and longitudinal polarization of
$J/\psi$ at the matrix element level. The calculations of spin
density matrix for the $Q\bar{Q}\lbrack
^3S_1^1,^3S_1^8,^1S_0^8\rbrack$ are as what have been done in other
similar processes~\cite{Gong:2008ft}.

In handling the processes in the COM, there are two aspects
which are different from the color-singlet case. The first is
is that in process (\ref{eqn:r4com}) $\gamma$ has
chance to collinear with quark or anti-quark $q(\bar{q})$ in
final states in some region of the phase space. This infrared
divergence will cancel the infrared divergence in the QED correction of $pp
\to J/\psi+g$. Because we observe the photon in the final states, it
means that we have to impose a cut on this process to
isolate a photon from the quark jet~\cite{Frixione:1998jh}
\bea
p_t^i\le p_t^{\gamma}\frac{1-\cos R_{\gamma_i}}{1-\cos
\delta_0}~~~~for~~~~R_{\gamma_i} < \delta_0.
\eea
The definitions of
the $p_t^i$, $p_t^\gamma$, $\cos R_{\gamma_i}$ and the
$\delta_0$ can be find in Ref.~\cite{Frixione:1998jh}. Here we
set $\delta_0=$0.7. For the consideration on the experimental
measurement, we also set cut-off on the transverse momentum of the
photon($p_t^{\gamma}$). Therefore the numerical results will rely
heavily on this condition.
The second is that the color-octet P-wave parts have additional infrared
divergence which will be factorized into the LDMEs by using
the same method as in Ref.~\cite{Wang:2012tz}.
In the calculation of real process $Q\bar{Q}\lbrack ^3P_J^8\rbrack + \gamma +g$
hadroproduction, there is a soft divergence related to $Q\bar{Q}$ pair radiating
the soft gluon and it can be factorized as an amplitude square of $Q\bar{Q}\lbrack ^3S_1^8,~^3S_1^1\rbrack +\gamma$
hadroproduction times a soft factor which contain soft divergence. This divergence can be absorbed into
the redefinition of the $Q\bar{Q}\lbrack ^3S_1^8,~^3S_1^1\rbrack$ LDMEs at NLO and there are finite parts being left.
Therefore, except the direct calculation of $Q\bar{Q}\lbrack ^3P_J^8\rbrack$ state we also have to take into account the
contribution from the left parts, which we call the q-term parts.

After generating the Fortran codes of these processes individually by
using the Feynman Diagram Calculation (FDC) package~\cite{Wang:2004du} ,
we had checked the cancelation of infrared and ultraviolet
divergence, the gauge invariance and the cut-independence
respectively. Because of the complexity of the analytic expressions
we use the quadruple precision program in some of the calculation to
avoid the numerical instability.

To obtain the numerical results, we choose the following
parameters and cut conditions. The charm quark mass $m_c$ is
set as 1.5GeV and will vary from 1.4GeV to 1.6GeV to estimate the
related uncertainty. The renormalization and factorization scales are set to
$\mu_r=\mu_f=\mu_0=\sqrt{(2m_c)^2+p_t^2}$ and it will vary from
$\mu_0/2$ to $2\mu_0$ to estimate the uncertainties. The NRQCD scale
$\mu_{\Lambda}$ is chosen as $m_c$. As for the experimental
conditions, we use $\sqrt{s}=7,8,14$TeV at the LHC,
the rapidity cuts $|y_{J/\psi,\gamma}|\le3$, or pseudo-rapidity cut $|\eta_{\gamma}|\le1.45$  and
$p_t^{\gamma}<1.5,3,5,15$GeV cuts. The fine structure
constant is chosen as $\frac{1}{128}$. The CTEQ6L1 and the CTEQ6M
PDFs and the corresponding $\alpha_s$ running formula are used to
calculate the LO and the NLO numerical results~\cite{Pumplin:2002vw}.

The involved LDMEs were extracted at the NLO by different group with different
consideration~\cite{Butenschoen:2011yh,Ma:2010yw,Chao:2012iv,Gong:2012ug}.
In Ref.~\cite{Butenschoen:2012qr} the authors investigated these LDME sets
and concluded that the universality of LDMEs is challenged.
The two LDME sets in Ref.~\cite{Ma:2010yw,Chao:2012iv}
are from the same group, and we use their former results on the combination of LDMEs in
Ref.~\cite{Ma:2010yw} to estimate numerical results since the feeddown contribution from $\chi_c$ and $\psi'$
had been considered there, which could
affect the theoretical prediction largely as discussed in Ref.~\cite{Gong:2012ug}.
We list these LDME sets in table \ref{tab:LDME}.
For the LDMEs in Reference~\cite{Ma:2010yw}, only the combinations of them,
$M^{J/\psi}_{0,r_0}$ and $M^{J/\psi}_{1,r_1}$, are given as
\bea
M^{J/\psi}_{0,r_0}=\langle
O^{J/\psi}(^1S_0^8)\rangle+\frac{r_0}{m^2_c}\langle
O^{J/\psi}(^3P_0^8)\rangle,\label{eqn:M0}\\
M^{J/\psi}_{1,r_1}=\langle
O^{J/\psi}(^3S_1^8)\rangle+\frac{r_1}{m^2_c}\langle
O^{J/\psi}(^3P_0^8)\rangle, \label{eqn:M1}
\eea
where $r_0$=3.9, $r_1$=-0.56, $M^{J/\psi}_{0,r_0}=0.074$
and $M^{J/\psi}_{1,r_1}=0.0005$.
With requiring the LDMEs to be positive we set the three individual
color-octet LDMEs from the above combinations under two conditions
in table~\ref{tab:LDME}, which we will refer to as "Ma extension"
in the following parts.
\begin{table}
\caption[]{The NRQCD LDMEs $\langle O^{J/\psi}(n)\rangle $ extracted
by three groups in
Ref.~\cite{Butenschoen:2011yh,Ma:2010yw,Chao:2012iv,Gong:2012ug}
at the NLO with $\langle O^{J/\psi}(^3S_1^1)\rangle =$1.32 (1.16) GeV$^3$
used in Ref.~\cite{Butenschoen:2011yh} (in the others).
The NRQCD LDMEs in Ma extension1
and extension2 are determined from the combination extracted in
Ref.~\cite{Ma:2010yw}.
}\label{tab:LDME}
%\begin{center}
\renewcommand{\arraystretch}{1.5}
\[
\begin{array}{|c|c|c|c|}
\hline \hline n &^1S_0^8,\mathrm{GeV}^3& ^3S_1^8,\mathrm{GeV}^3& ^3P_0^8,\mathrm{GeV}^5 \\
\hline \textrm{Butenschoen\cite{Butenschoen:2011yh}}&0.0497&0.0022&-0.0161\\
\hline \textrm{Gong\cite{Gong:2012ug}}&0.097&-0.0046&-0.0214\\
\hline \textrm{Chao\cite{Chao:2012iv}}&0.089&0.0030&0.0126\\
\hline \textrm{Ma~extension1}&0.074&0.0005&0\\
\hline \textrm{Ma~extension2}&0&0.011&0.019\\
\hline \hline
\end{array}
\]
\renewcommand{\arraystretch}{1.0}
%\end{center}
\end{table}

\begin{figure}
\includegraphics[scale=0.35]{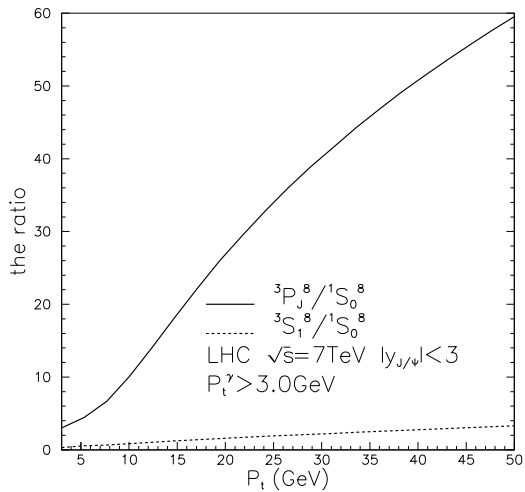}
\caption{\label{fig:pr7} The ratio of $d\sigma(^3P_J^8)/d\sigma(^1S_0^8)$ and
$d\sigma(^3S_1^8)/d\sigma(^1S_0^8)$ as functions of $p_t$.}
\end{figure}

It is shown in Fig.\ref{fig:pr7} that the color-octet $^3P_J^8$
state, just like $^3S_1^8$ and $^1S_0^8$ state, gives a positive
short distance coefficient in all $p_t$ region in contrast to
$J/\psi$ inclusive hadroproduction at QCD NLO case where the
color-octet $^3P_J^8$ state gives negative short distance
coefficient in large $p_t$ region. The difference makes the $p_t$
distribution of $J/\psi+\gamma$ production rate an observable which
can break the linear correlation in previous fit.

\begin{figure*}
\begin{center}
\includegraphics[scale=0.29]{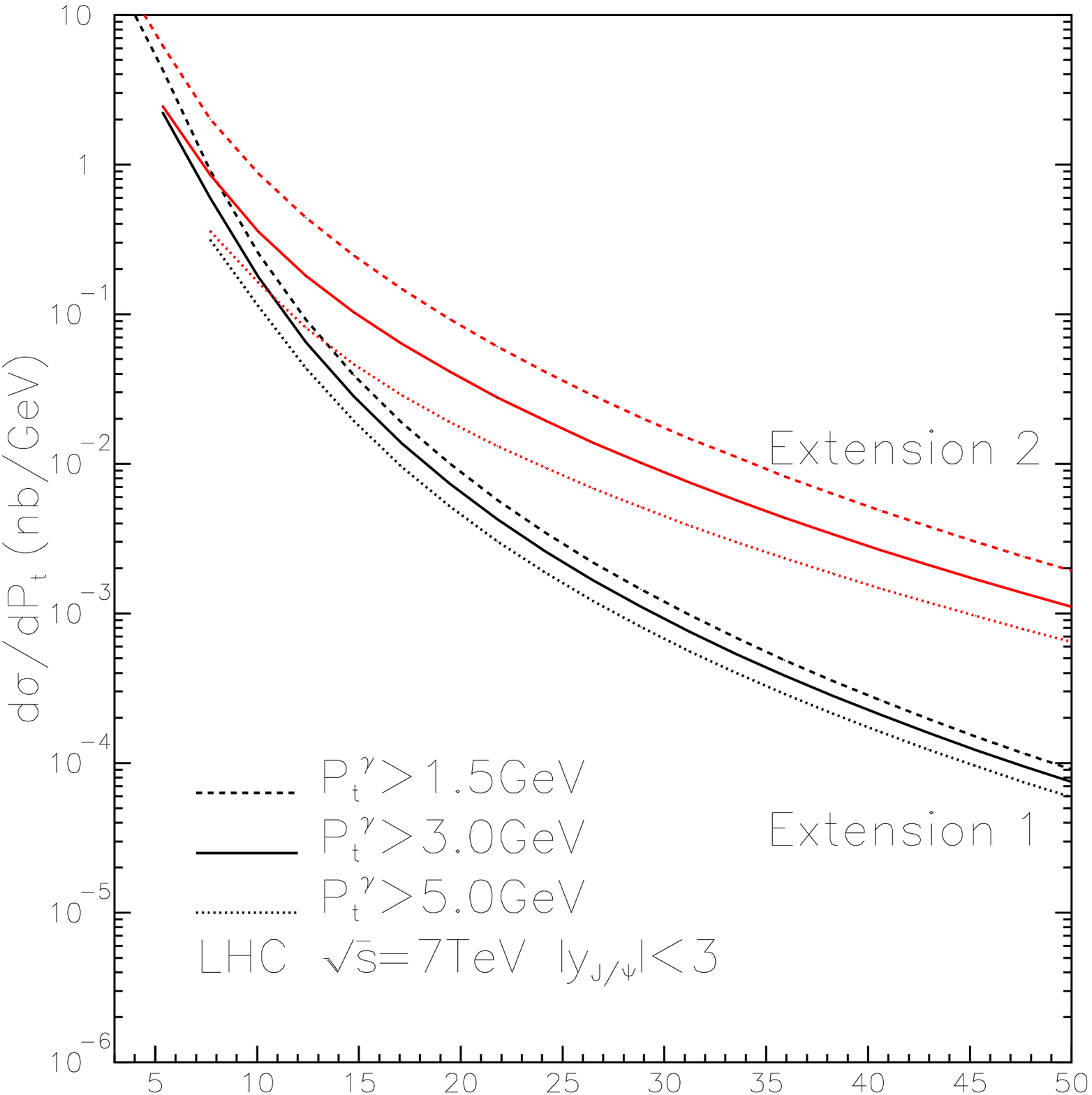}
\includegraphics[scale=0.29]{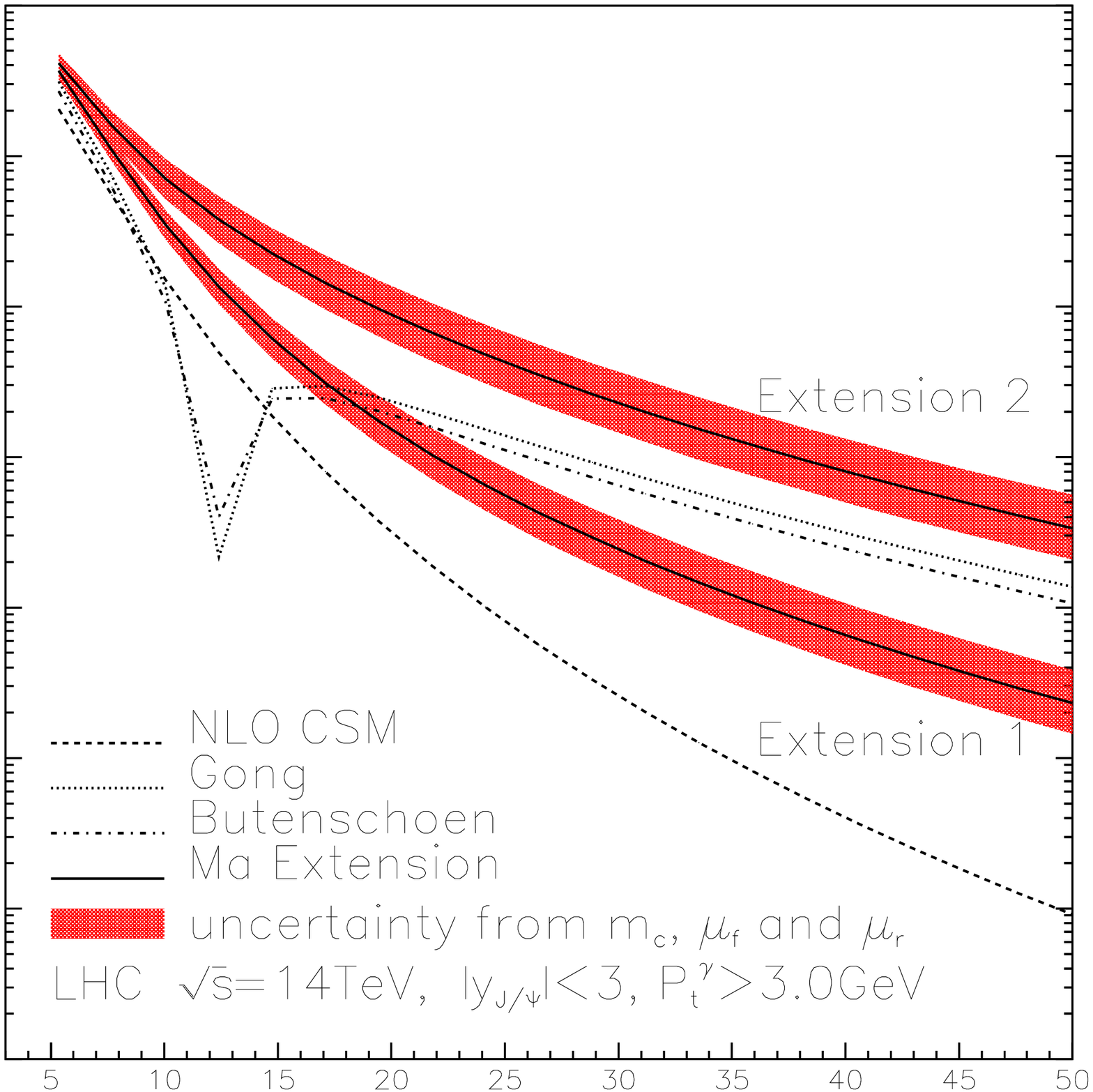}
\includegraphics[scale=0.29]{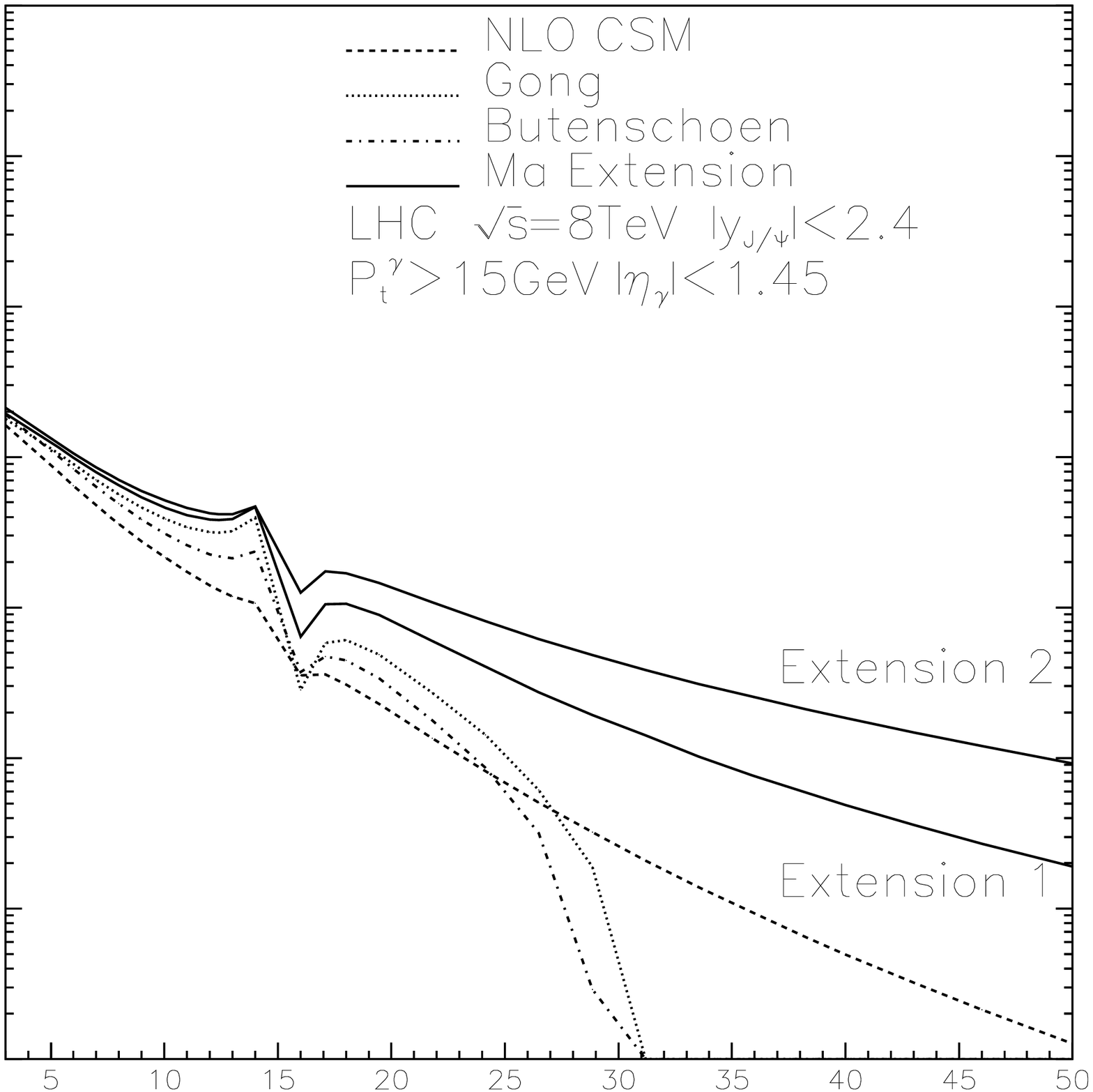}\\
\includegraphics[scale=0.29]{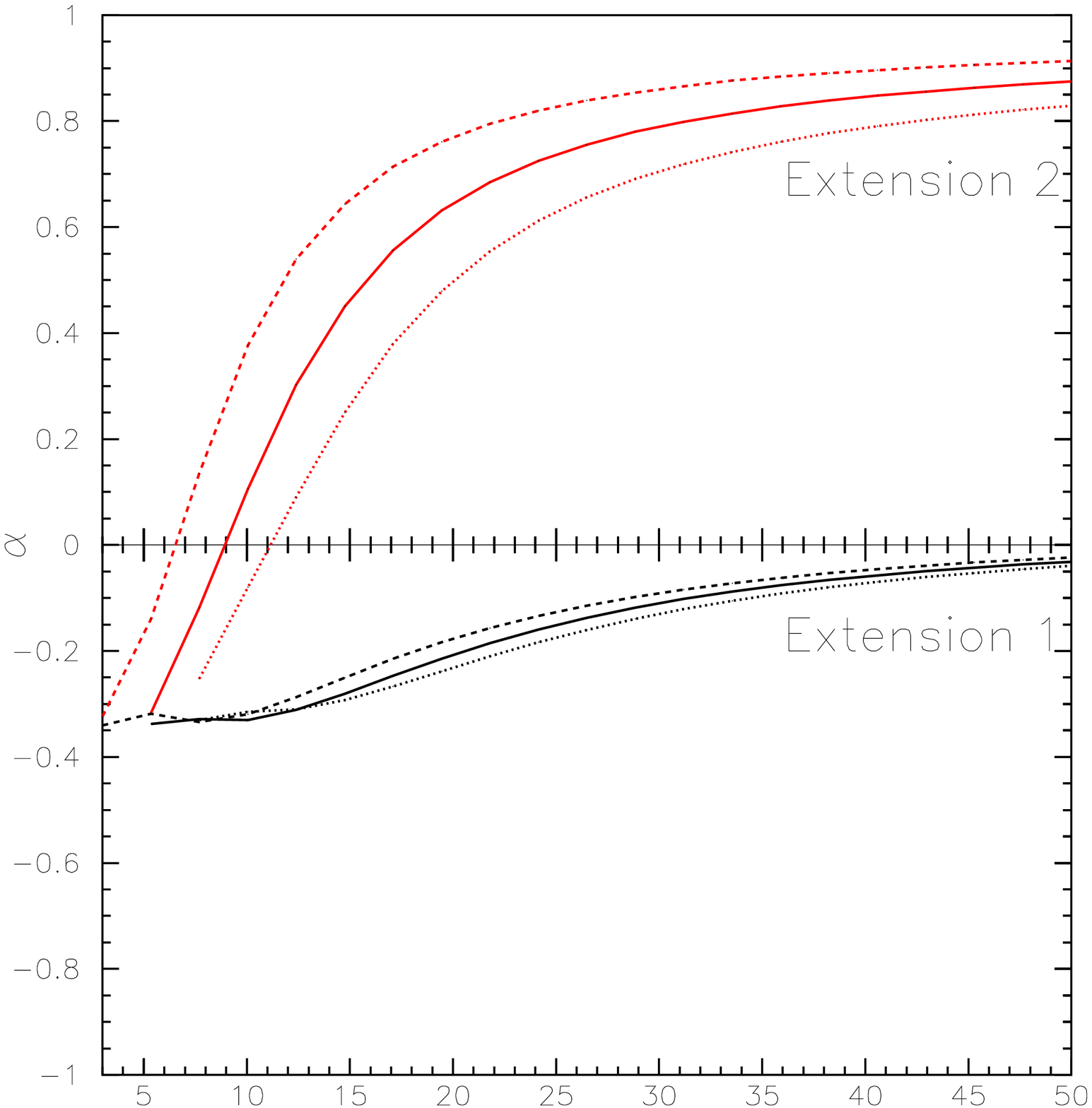}
\includegraphics[scale=0.29]{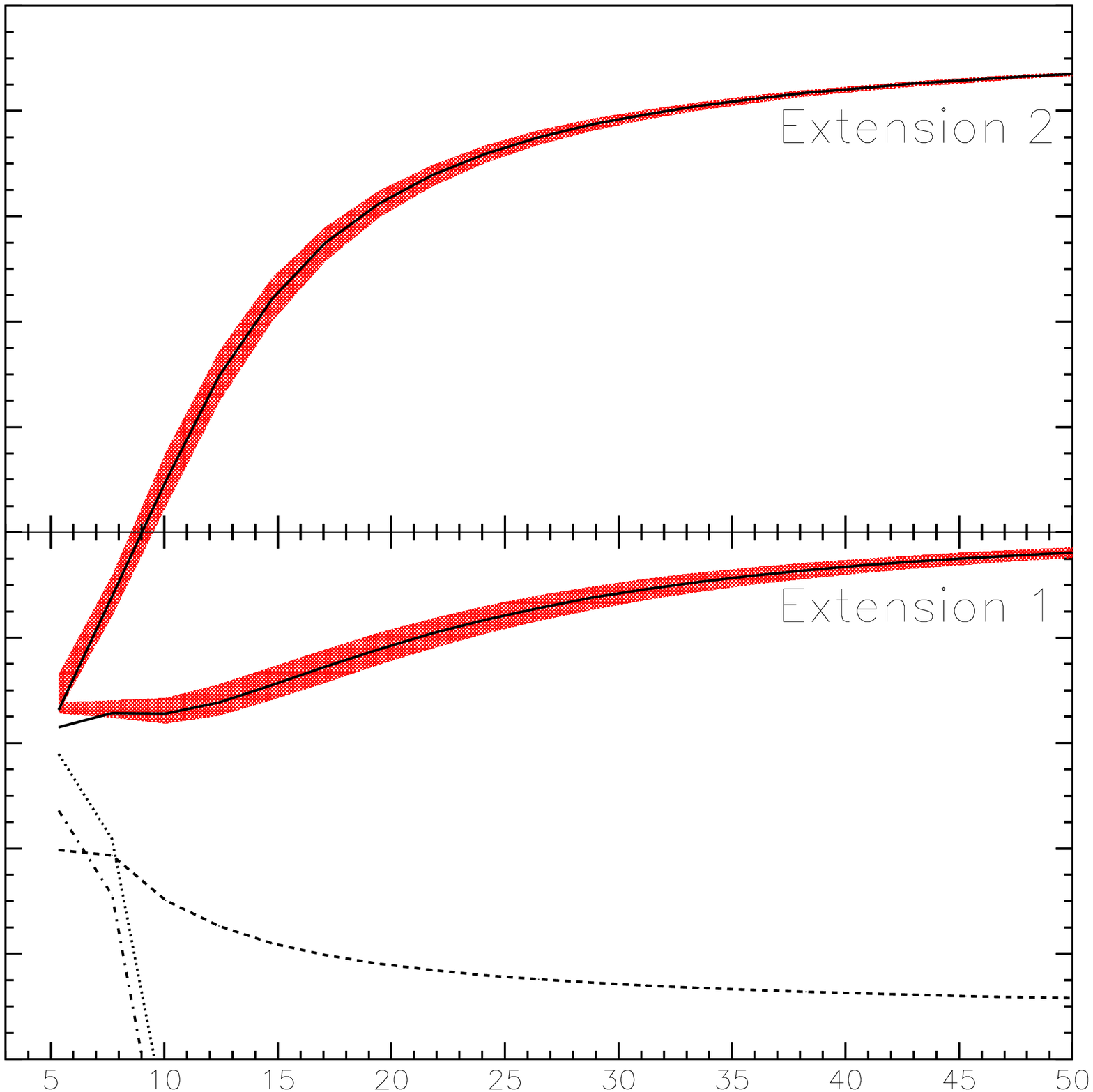}
\includegraphics[scale=0.29]{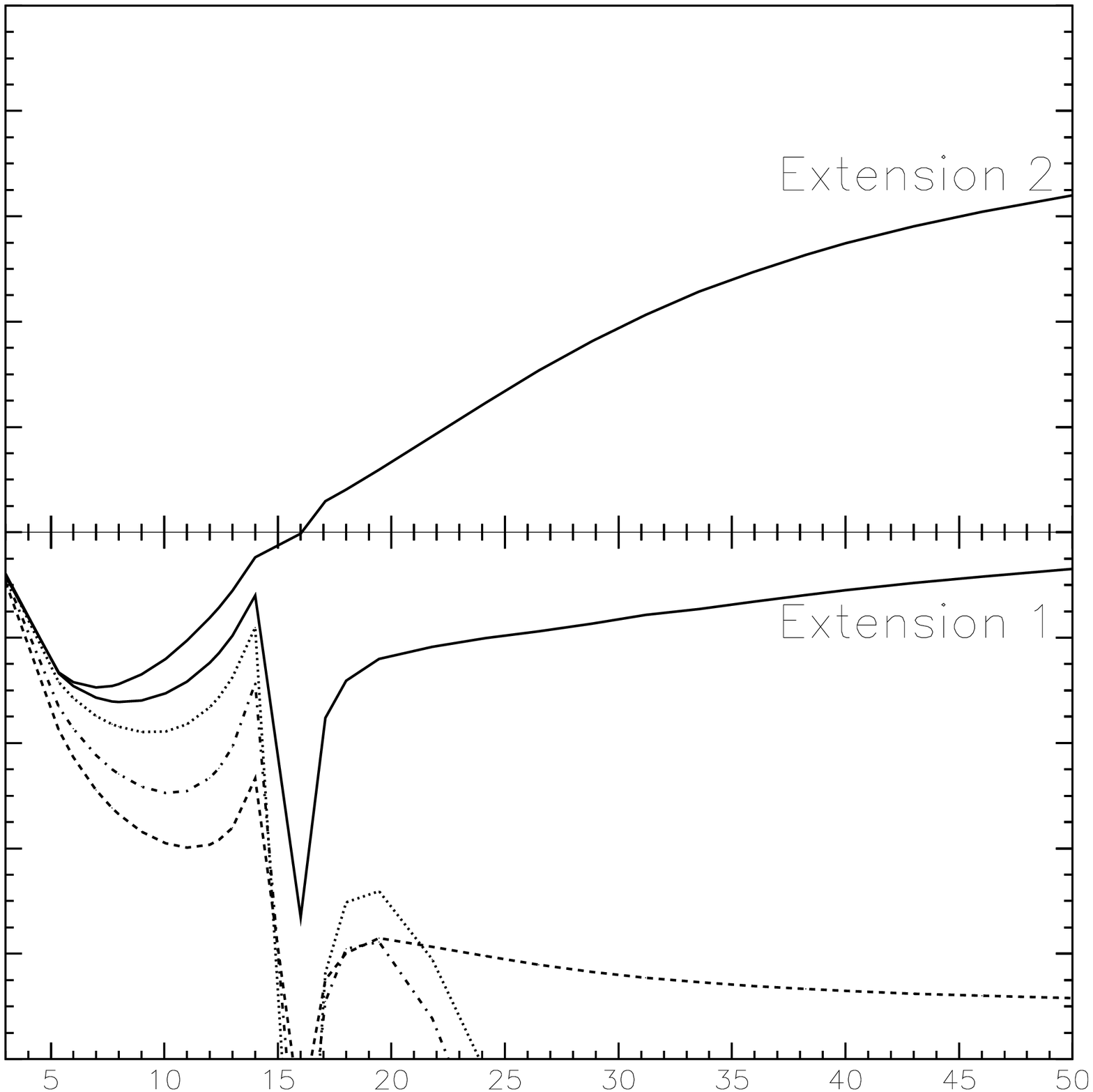}
\caption{\label{fig:ptdis} The $p_t(J/\psi)$
distributions for $J/\psi+\gamma$ production (upper parts) and polarization
(lower parts) with different conditions. Figures in the
same column are of the same conditions and line types.
The shaded band in the second column represent the uncertainty
from variation of $\mu_f$ \& $\mu_r$ and $m_c$.
The third column shows results with $p_t^{\gamma}>15$GeV and different LDME sets.}
\end{center}
\end{figure*}

The first column of Fig.\ref{fig:ptdis} presents the results at $\sqrt{s}=7$TeV.
In order to investigate the dependence on $p_t^{\gamma}$ cut, we plot
the $p_t$ distribution of production rate and
polarization observable $\alpha$ with different $p_t^{\gamma}$ cuts and two
sets of the LDMEs in Ma extension. The plots show that the dependence
on $p_t^{\gamma}$ cut decreases with the increase of $p_t(J/\psi)$ and different
$p_t^{\gamma}$ cuts make the $p_t$ distribution shift parallel in contrast to the
color-singlet channel in Ref.~\cite{Li:2008ym} where the $p_t$ distribution of $J/\psi$
production rate and $\alpha$ are insensitive to the the $p_t^{\gamma}$ cut in the same $p_t$
region.  The reason is that $^3P_J^8$ and $^1S_0^8$ channel in the COM are sensitive to the cuts.

In the second column in Fig.\ref{fig:ptdis}, we give results at $\sqrt{s}=14$TeV.
The shaded band represents the uncertainties estimated by varying the $m_c$,
the renormalization scale $\mu_r$ and factorization scale $\mu_f$.
The plots show that the uncertainties of production rate become larger and
that of $\alpha$ become smaller as $p_t$ increasing.  The COM contribution on production rate dominant
over that of CSM and are about 2 orders larger than the color-singlet ones at $p_t=50$GeV.
We also plot the $p_t$ distribution with the LDME sets in Ref.~\cite{Butenschoen:2011yh}
and \cite{Gong:2012ug},  the absolute value of the numerical results are used in the $p_t$ distribution
of production rate since they become negative when $p_t>13$GeV,
and $\alpha$s in both cases are out of physical region when $p_t>10GeV$.

From the results at the first and second columns of Fig.\ref{fig:ptdis}, we know that
the $p_t$ distribution of $J/\psi+\gamma$ hadroproduction rate is good observable to
distinguish different LDME sets. Is it measurable or not at the 8GeV LHC with present 23fb$^{-1}$
integrated luminosity?  To suppress the background efficiently, $p_t^{\gamma}>15$GeV cut on photon is applied,
together with $|y_{J/\psi}|<$2.4 and $|\eta_{\gamma}|<$1.45 for photon reconstruction efficiency consideration.
The plots in the third column of Fig.\ref{fig:ptdis} show that the $p_t$
distributions of $J/\psi$ production rate in the COM with Ma extension LDME sets are
about 10(100) times larger than that in the CSM.  The other two LDME sets give the
positive predictions in lower $p_t$ region and negative ones when $p_t$ is larger than 31 GeV.
When $p_t$ is larger than 20GeV, the results show many differences on the $J/\psi$ polarization
predictions $\alpha$ with the CSM (COM) mechanism and three LDME sets. It is mentionable
that only real processes at QCD NLO contribute when $p_t^{J/\psi}<15$GeV.

In summary, we present the study on associate $J/\psi+\gamma$ hadroproduction at the NLO
with full COM contribution at the LHC.  Our numerical results show that
the contribution from color-octet channels enhances the
differential cross section about 2 order in the large $p_t$ region.
As for the $J/\psi$ polarization, the color-octet contribution
changes it from longitudinal one to transverse one.
From all the plots in Fig.\ref{fig:ptdis},
it is manifestly that the most important uncertainty comes from the the variation of LDMEs.
The LDME sets of Butenschoen and Gong  lead to the unphysical
$p_t$ distribution of production rate (negative) and polarization observable $\alpha$
(out of range -1 to 1) at large $p_t$ range, while the
LDME set of Ma extension gives physical ones at all the $p_t$ range.
Even within Ma extension, the $p_t$ distributions of production rate are of huge
difference ( 10 times at $p_t=50$GeV) between the extension2 and extension1.
The polarization observable $\alpha$  changes from slightly
longitudinal in extension1 to the transverse in extension2.

In conclusion, the theoretical predictions are sensitive to the
LDMEs heavily and can break the linear correlation in previous fit.
To obtain an experimental measurable observable at the 8GeV LHC with
present 23fb$^{-1}$ integrated luminosity, $p_t^{\gamma}>15$GeV cut
on the observed photon is applied to efficiently suppress the
background and $|\eta_{\gamma}|<$1.45 is used.  With these conditions, 
the photon reconstruction efficiency is larger than $0.7$ and we use $0.7$ in the following estimation, 
$Br(J/\psi\rightarrow\mu^++\mu^-)=0.05$ is also used to represent reconstruction
of $J/\psi$ from the observed $\mu^++\mu^-$ pair.  Then the plots in the
third column of Fig.\ref{fig:ptdis} indicates that 800-1600 events
at $p_t=17$Gev and 16-80 events at $p_t=50$GeV could be
reconstructed from the sample data. Therefore, the $p_t$
distribution of production rate is experimental measurable with
present data sample collected at $8$TeV LHC. All the color-octet
LDMEs in $J/\psi$ production could be fixed sensitively by including
this proposed measurement and our calculation.

We thank Dr. Bin Gong, Hong-Fei zhang and Lu-Ping Wan for helpful
discussions. Rong Li also thanks Qiang Li for the discussion on the
isolation of photon. This work is supported by the National Natural
Science Foundation of China (No.~11105152, 10935012, and No. 11005137),
DFG and NSFC (CRC110), and by CAS under Project No. INFO-115-B01.

%\bibliography{gg-Jpsigamma}% Produces the bibliography via BibTeX.

\end{document}